\documentclass[twocolumn,epj,nopacs]{svjour}
\usepackage{graphicx}
\usepackage{dcolumn}
\usepackage{bm}
\usepackage{amssymb}
\usepackage{multirow}
\usepackage{epsfig}
\usepackage{subfigure}
\usepackage{atlasphysics}
\usepackage{epstopdf}
\usepackage{textpos}
\usepackage{cite}
\usepackage{hyperref}
\usepackage{stfloats}

\usepackage{preprintcover} 
\PreprintCoverPaperTitle{Search for Lepton Flavour Violation in the $e\mu$ Continuum with the ATLAS Detector in $\sqrt{s} = 7$ TeV $pp$ Collisions at the LHC} 
\PreprintIdNumber{CERN-PH-EP-2012-108} 
\PreprintCoverAbstract{This paper presents a search for the $t$-channel exchange of an $R$-parity violating scalar top quark ($\tilde{t}$) 
in the $e^{\pm}\mu^{\mp}$ continuum using 2.1~fb$^{-1}$ of data collected by the ATLAS detector 
in $\sqrt{s}=7$ TeV $pp$ collisions at the Large Hadron Collider. 
Data are found to be consistent with the expectation from the Standard Model backgrounds. 
Limits on $R$-parity-violating couplings at 95\% C.L. are calculated as a function of the scalar top mass ($m_{\tilde{t}}$).
The upper limits on the production cross section for $pp~\rightarrow~e\mu X$, through the $t$-channel exchange of a scalar top quark, ranges from 170~fb for $m_{\tilde{t}}=95$~GeV to 30~fb for $m_{\tilde{t}}=1000$~GeV.
} 
\PreprintJournalName{European Physical Journal C}

\newcommand{\vmet}{\ensuremath{\vec{\slash\kern-.7emE}_{T}}}

\newcommand{\memu}{\ensuremath{m_{e\mu}}}
\newcommand{\mstop}{\ensuremath{m_{\tilde{t}}}}

\begin{document}

\title{Search for Lepton Flavour Violation in the $e\mu$ Continuum with the ATLAS Detector in $\sqrt{s} = 7$ TeV $pp$ Collisions at the LHC
}
\titlerunning{Search for Lepton Flavour Violation in the $e\mu$ Continuum}
\author{The ATLAS Collaboration}
\authorrunning{The ATLAS Collaboration}
\institute{CERN, 1211 Geneva 23, Switzerland \email{atlas.publications@cern.ch}}
\date{\today}

\abstract{
This paper presents a search for the $t$-channel exchange of an $R$-parity violating scalar top quark ($\tilde{t}$) 
in the $e^{\pm}\mu^{\mp}$ continuum using 2.1~fb$^{-1}$ of data collected by the ATLAS detector 
in $\sqrt{s}=7$ TeV $pp$ collisions at the Large Hadron Collider. 
Data are found to be consistent with the expectation from the Standard Model backgrounds. 
Limits on $R$-parity-violating couplings at 95\% C.L. are calculated as a function of the scalar top mass ($m_{\tilde{t}}$).
The upper limits on the production cross section for $pp~\rightarrow~e\mu X$, through the $t$-channel exchange of a scalar top quark, ranges from 170~fb for $m_{\tilde{t}}=95$~GeV to 30~fb for $m_{\tilde{t}}=1000$~GeV.
}

\maketitle
\section{Introduction}
In the Standard Model (SM), direct production of $e^{\pm} \mu^{\mp}$ ($e\mu$) pairs is forbidden in $pp$ collisions due to 
lepton flavour conservation.  However, in many extensions of the SM, lepton flavour violation (LFV) is permitted. 
In particular, $R$-parity-violating (RPV) supersymmetric (SUSY) models, LFV leptoquarks, and models with additional gauge symmetry allow LFV.  
Previous searches by the CDF, D0, and ATLAS Collaborations~\cite{RPVCDF1,RPVCDF2,RPVCDF,RPVD01,RPVD0,emu_paper1,emu_paper2} 
have focused on resonant production of a heavy neutral particle which decays into an $e\mu$ pair and have set limits on these models.  
In addition to resonant $e \mu$ production, RPV SUSY models also allow for LFV interactions through the $t$-channel exchange of a scalar quark. The corresponding Lagrangian term for these RPV processes~\cite{rpv_sneutrino} is
$ {\cal{W}} = - \lambda_{ijk}^{\prime} \tilde{u}_{j} \bar{d}_{k} \ell_{i}$, 
where $\tilde{u}$ denotes the up-type squark field, $d$ is the down-type quark field, 
$\ell$ represents the lepton field,  and $\lambda'$  is the coupling at the production vertex. 
The indices $i,j,k$ refer to fermion generations.
This superpotential couples an up-type squark to a down-type quark and a lepton, allowing for production of 
$e\mu$ pairs through the $t$-channel exchange of an up-type squark.  
This paper presents a search for this process in the $e\mu$ continuum using 
2.1~fb$^{-1}$ of $pp$ collision data at $\sqrt{s} = 7$ TeV collected by the ATLAS detector at the Large Hadron Collider (LHC). 

\begin{figure}[htb]
\begin{center}
  \includegraphics[width=0.4\textwidth]{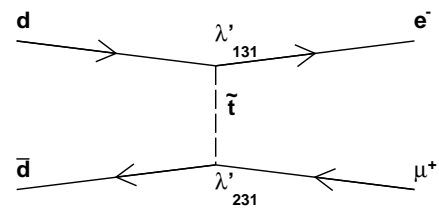}
\end{center}
\caption{The Feynman diagram for $d \bar{d} \rightarrow e^- \mu^+$ production through the $t$-channel exchange of a scalar top quark.}
\label{fig:diagram}
\end{figure}

The cross section for this process is expected to be dominated by the lightest up-type squark, which is taken to be the scalar top 
quark ($\tilde{t}$) in this analysis.  The Feynman diagram for the dominant process, $d\bar{d} \rightarrow e^-\mu^+$ 
through the $t$-channel exchange of a $\tilde{t}$, is shown in Fig.~\ref{fig:diagram}. The leading-order (LO) 
partonic differential cross section is calculated as
 $d\hat{\sigma} / d\hat{t}  = |\lambda_{131}^{\prime}\lambda_{231}^{\prime}|^2 \hat{t}^2 / [64 N_{\rm{c}} \pi \hat{s}^2 \left( \hat{t} - m_{\tilde{t}}^2 \right)^2]$,
where $\hat{s}$ and $\hat{t}$ are the usual Mandelstam variables in the $d\bar{d}$ centre-of-mass frame, $N_{\rm{c}}=3$ is 
the colour factor, $m_{\tilde{t}}$ is the scalar top mass, and $\lambda^{\prime}_{131}$ ($\lambda^{\prime}_{231}$) is the coupling for 
the vertex $d\tilde{t}e^-$ ($\mu^+ \tilde{t}\bar{d}$).  
The process where the final state leptons have opposite charges to those in Fig.~\ref{fig:diagram} has the same cross section. 
Diagrams with the $d$ and $\bar{d}$ independently replaced by $s$ and $\bar{s}$ quarks are also allowed. 
The form of the cross section for these diagrams is the same, but the indices on the $\lambda^{\prime}$ couplings are different.
In the case of 
$s \bar{s}  \rightarrow \mu^{\pm} e^{\mp}$, the cross section depends on $|\lambda_{132}^{\prime}\lambda_{232}^{\prime}|$.  
For $d \bar{s} \rightarrow \mu^{+}  e^{-} $ and $s \bar{d} \rightarrow \mu^{-} e^{+} $, the cross section depends 
on $|\lambda_{131}^{\prime}\lambda_{232}^{\prime}|$.  Lastly, diagrams with $s \bar{d} \rightarrow \mu^{+} e^{-} $ and $d \bar{s}\rightarrow \mu^{-} e^{+} $ depend on $|\lambda_{231}^{\prime}\lambda_{132}^{\prime}|$.

     Strong limits on RPV couplings have been obtained from low-energy searches~\cite{rpv-phenom,rpv-phenom2}, such as $\mu \to e \gamma$, $\mu - e$ conversion on nuclei and $Z \to e \mu$, where superparticles appear in the intermediate state, often in loops.  The presence of multiple interfering amplitudes makes the extraction of limits difficult, and it is usually assumed that a single product of couplings dominates.  The interference of different diagrams could weaken the limits on a specific product of couplings.  Also, these limits depend on unknown superparticle masses (including ones other than the scalar top), sometimes in a complex manner.  

The HERA experiments searched for an LFV leptoquark in the process $e p \to \mu X$~\cite{H1,Zeus1}.  These studies also place limits on a potential RPV scalar top.  At lower masses (less than about 300~GeV), there would be copious $s$-channel production, and placing limits on specific couplings depends on assumptions about the stop decays. At higher masses, the HERA searches are sensitive to $u$-channel exchange, which can be directly compared to this analysis.  The sensitivity of the measurement in this paper is slightly better than at HERA for masses above about 300~GeV. The HERA experiments also searched for scalar top production in both the RPV and gauge boson decay channels~\cite{Zeus2,H12}. Such searches assumed the RPV coupling involved in the scalar top production,
$\lambda^{\prime}_{131}$, to be dominant and cannot be directly compared with the results of this paper.

     Direct searches at hadron colliders and at HERA for lepton-flavour-conserving scalar leptoquarks~\cite{leptoQ1,leptoQ2,leptoQ3,leptoQ4,leptoQ5,leptoQ6,leptoQ7,leptoQ8,leptoQ9,leptoQ10} are also relevant to the search here.  The interpretation of such results as limits on a scalar top depends, as for the LFV leptoquarks, on the decay branching ratios to the leptons and quarks and hence on assumptions about the other possible decays. Present limits on such leptoquarks at the scalar top masses considered here do not preclude the signal sought in this analysis.

The limits on the couplings associated with the $d\bar{s}$ and $s\bar{d}$ processes are two orders of magnitude lower than those for the $d\bar{d}$ and $s\bar{s}$ 
couplings~\cite{rpv-phenom}.  Therefore dominance by same flavour quark scattering processes is assumed in this analysis.  As a result, the production cross section for 
$pp~\rightarrow~e\mu X$, due to the $t$-channel exchange of a scalar top quark, depends on $\lambda_{131}^{\prime}$, $\lambda_{231}^{\prime}$, 
$\lambda_{132}^{\prime}$, $\lambda_{232}^{\prime}$, and $m_{\tilde{t}}$.  

\section{Detector and Data Sample}
The \mbox{ATLAS} detector~\cite{atlas_detector} is a multi-purpose particle detector with a forward-backward symmetric 
cylindrical geometry and almost $4\pi$ coverage in solid angle~\cite{coordinate}.
The inner tracking detector (ID) covers $|\eta|<2.5$ in pseudorapidity $\eta$ and consists of a silicon
pixel detector, a silicon microstrip detector, and a transition 
radiation tracker. The ID is surrounded by a thin superconducting solenoid providing a
2~T magnetic field and by a hermetic calorimeter system, 
which provides three-dimensional reconstruction of particle showers up to $|\eta|=4.9$. 
The muon spectrometer (MS) is based on one barrel and two endcap air-core toroids, 
each consisting of eight superconducting coils arranged
symmetrically in azimuth around the calorimeter. 
Three layers of precision tracking stations, consisting of drift tubes
and cathode strip chambers, allow precise muon momentum measurement up
to $|\eta|=2.7$. Resistive plate and thin-gap chambers provide muon triggering capability up to $|\eta|=2.4$.

The $pp$ collision data used in this analysis were recorded between March and August 2011 at a centre-of-mass energy of 7 TeV. 
After applying data quality requirements, the total integrated luminosity of the dataset used in this analysis is 2.08 $\pm$ 0.08~fb$^{-1}$~\cite{lumi}. 
Events are required to satisfy one of the single-lepton ($e$ or $\mu$) triggers. 
For electrons, the threshold on the transverse energy (\et)  is 20~GeV or 22~GeV depending on run periods, and 
for muons the threshold on the transverse momentum (\pt) is 18~GeV. 

\section{Event Preselection}
The event preselection requires a primary vertex with at least three associated tracks with \pt~$> 0.5$~GeV and exactly one electron and one muon of opposite charge.  
Electron candidates are selected from clustered energy deposits in the electromagnetic 
calorimeter with an associated track reconstructed in the ID. They are required to have \et~$> 25$~GeV 
and to lie inside the pseudorapidity regions $|\eta| < 1.37$ or $1.52~<~|\eta|~<~2.47$.
Electrons are further required to satisfy a stringent set of identification requirements based on the calorimeter 
shower shape, track quality and track matching with the calorimeter energy cluster, referred to 
as `tight' in Ref.~\cite{electronMedium}. Muons are reconstructed by combining 
tracks in the ID and MS with \pt~$> 25$~GeV and $|\eta| < 2.4$.  Electrons are rejected if they are located within a cone of $\Delta R = \sqrt{(\Delta \eta)^2 + (\Delta \phi)^2} = 0.2$ around a muon, where $\Delta \eta$ and $\Delta \phi$ are the pseudorapidity and azimuthal opening angle difference between the electron and muon.

To suppress backgrounds from $W/Z$+jets and multijets, isolation requirements on tracks and calorimeter deposits are applied to the leptons. The scalar sum of the transverse momenta of tracks within a cone of $\Delta R = 0.2$ around the lepton must be less than 10\% of the lepton's \pt.  Similarly, the transverse energy in the calorimeter within a cone of $\Delta R = 0.2$ around the lepton are required to be less than 15\% of the lepton's transverse energy.  Corrections are applied to account for energy leakage and energy deposition inside the isolation cone due to additional $pp$ collisions.

Jets are reconstructed from calibrated clusters using
the anti-$k_t$ algorithm~\cite{antiKtTopo} with a radius parameter of 0.4.  Jet energies are calibrated using \et- and $\eta$-dependent correction factors based on Monte Carlo (MC) simulation and validated by test beam and collision data studies~\cite{jetscale}.  Only jets with $\pt>30$~GeV and $|\eta|<2.5$ are considered. 
If such a jet and an electron lie within $\Delta R=0.2$ of each other, the jet is discarded. 

The measurement of missing transverse momentum~\cite{met} (\met) is based on the transverse momenta of the electron and muon candidates, all jets, and all energy clusters with $|\eta| < 4.5$ not associated to such objects.

\section{Background and Simulation}
The SM processes that can produce an $e\mu$ signature are predominantly $t\bar{t}$,
$Z/\gamma^* \rightarrow \tau \tau$, diboson, single top, $W/Z$+jets, $W/Z$+$\gamma$  and multijet events. 
All of these processes, except $W/Z+$jets and multijet production,
are estimated using Monte Carlo samples generated at $\sqrt{s}=7$ TeV followed by a detailed {\sc geant4}-based~\cite{geant} simulation of the ATLAS detector~\cite{simulation}.  
To improve the agreement between data and simulation, selection efficiencies are measured in 
both data and simulation, and correction factors are applied to the simulation. 
Furthermore, the simulation is tuned to reproduce the calorimeter energy and the muon momentum scale and resolution.
Top production is generated with {\sc mc@nlo}~\cite{mcnlo} for $t\bar{t}$ and single top, the Drell-Yan process is generated with {\sc pythia}~\cite{pythia}, and the diboson processes are generated with {\sc herwig}~\cite{herwig}.
The $W/Z+\gamma$ background comes from the $W(\rightarrow \mu\nu)\gamma$ and $Z(\rightarrow \mu\mu)\gamma$ processes, which is estimated using events generated with {\sc madgraph}~\cite{madgraph}. The simulation samples are normalized to 
cross sections with higher-order corrections applied. 

The $\tilde{t}$ signal samples are produced with the {\sc pythia} event generator~\cite{pythia} 
with $|\lambda_{131}^{\prime} \lambda_{231}^{\prime}| = |\lambda_{132}^{\prime}\lambda_{232}^{\prime}| = 0.05$ 
and the value of $m_{\tilde{t}}$ is varied from 95~GeV, which is the most stringent limit from previous experiments~\cite{PDG}, to 1000~GeV.  The central CTEQ6L1~\cite{cteq6} parton distribution function (PDF) 
set is used. The LO cross section is 580~fb for $m_{\tilde{t}}=95$~GeV and 0.33~fb for $m_{\tilde{t}}=1000$~GeV.

\section{Data Analysis}
The production of $W/Z$+jets and multijets can give rise to backgrounds due to jets misidentified as leptons 
or non-prompt leptons from heavy-quark decays in jets. 
These sources are referred to as fake background and are estimated from data.
A looser lepton quality selection (called `loose' lepton here) is defined for each lepton type in addition to the default tight quality selection. For loose muons, both the calorimeter and the track isolation requirements are removed. 
For loose electrons, the `loose' electron identification criteria as defined in Ref.~\cite{electronMedium} are used and the 
isolation requirements are also removed.  The fake background is determined by weighting the events in the loose lepton sample by the likelihood that the event came from processes with at least one misidentified or non-prompt lepton.  These weights are obtained by solving a $4 \times 4$ matrix equation, 
constructed from the \et-~or \pt-dependent probabilities for a prompt or fake/non-prompt lepton that passes the loose lepton requirement to also pass the tight lepton requirement.  More details about the $4 \times4$ matrix method are given in Ref.~\cite{emu_paper2}. 

\begin{figure*}[htb]
\begin{center}
  \includegraphics[width=0.49\textwidth]{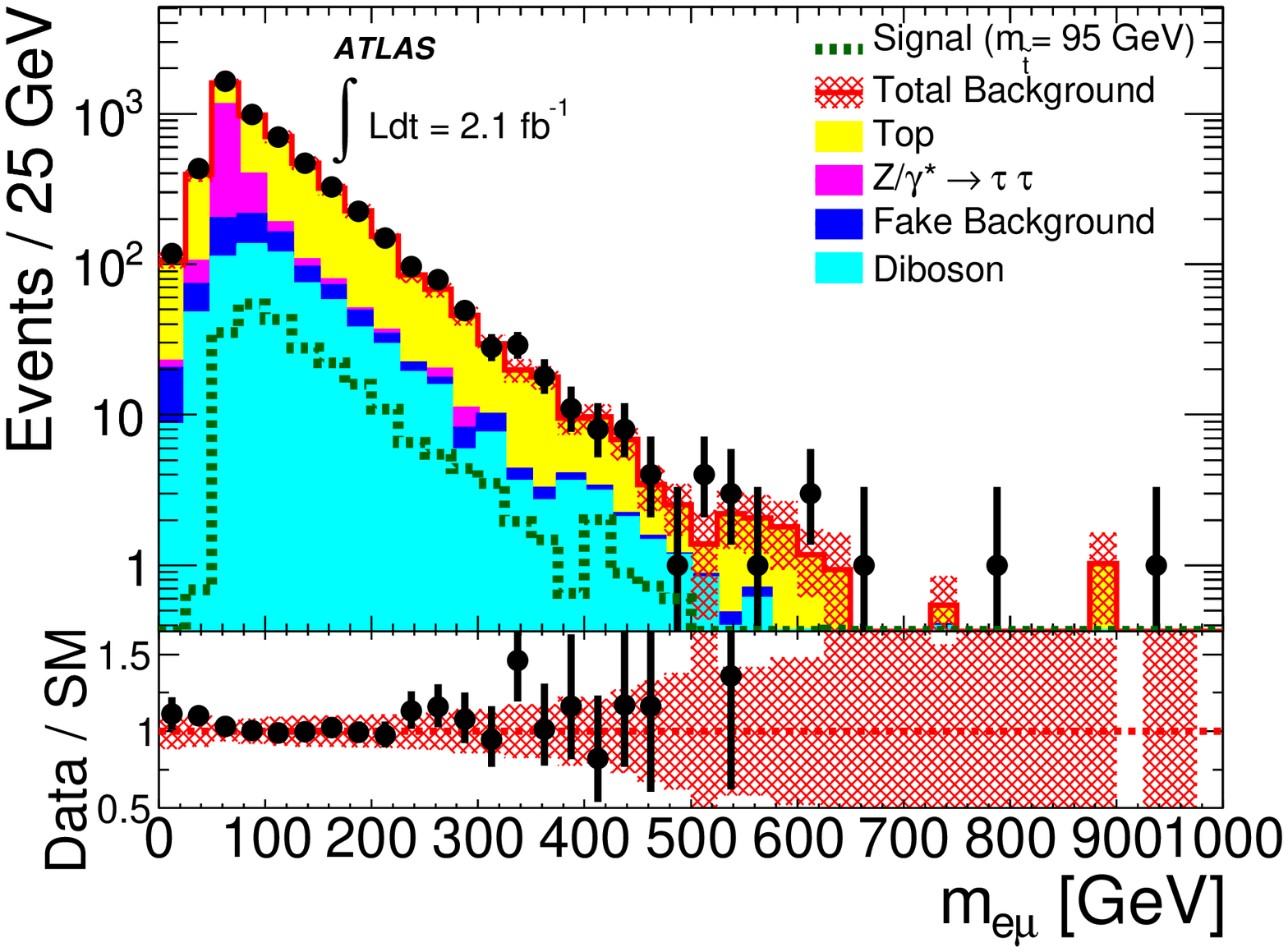}
  \includegraphics[width=0.49\textwidth]{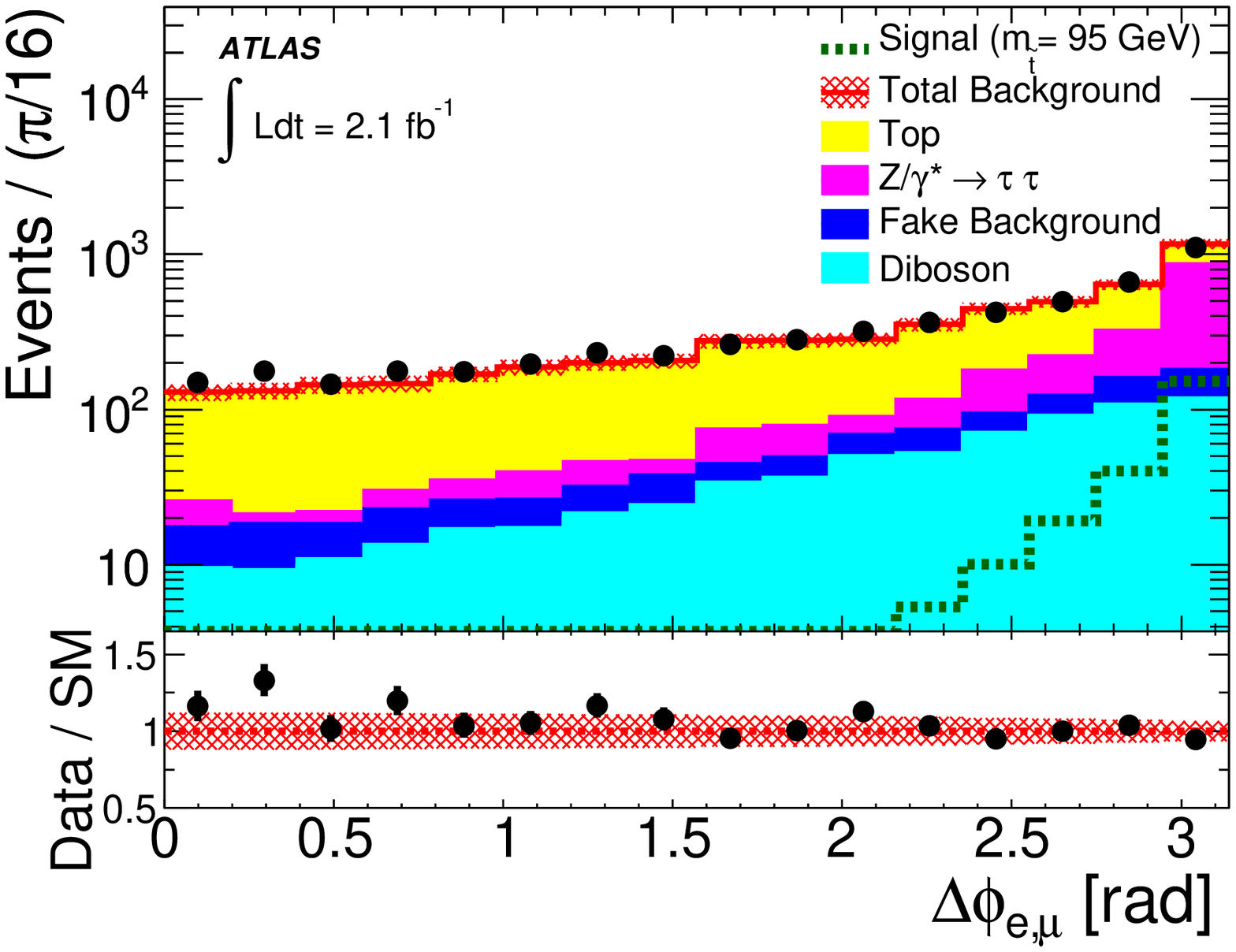}
  \includegraphics[width=0.49\textwidth]{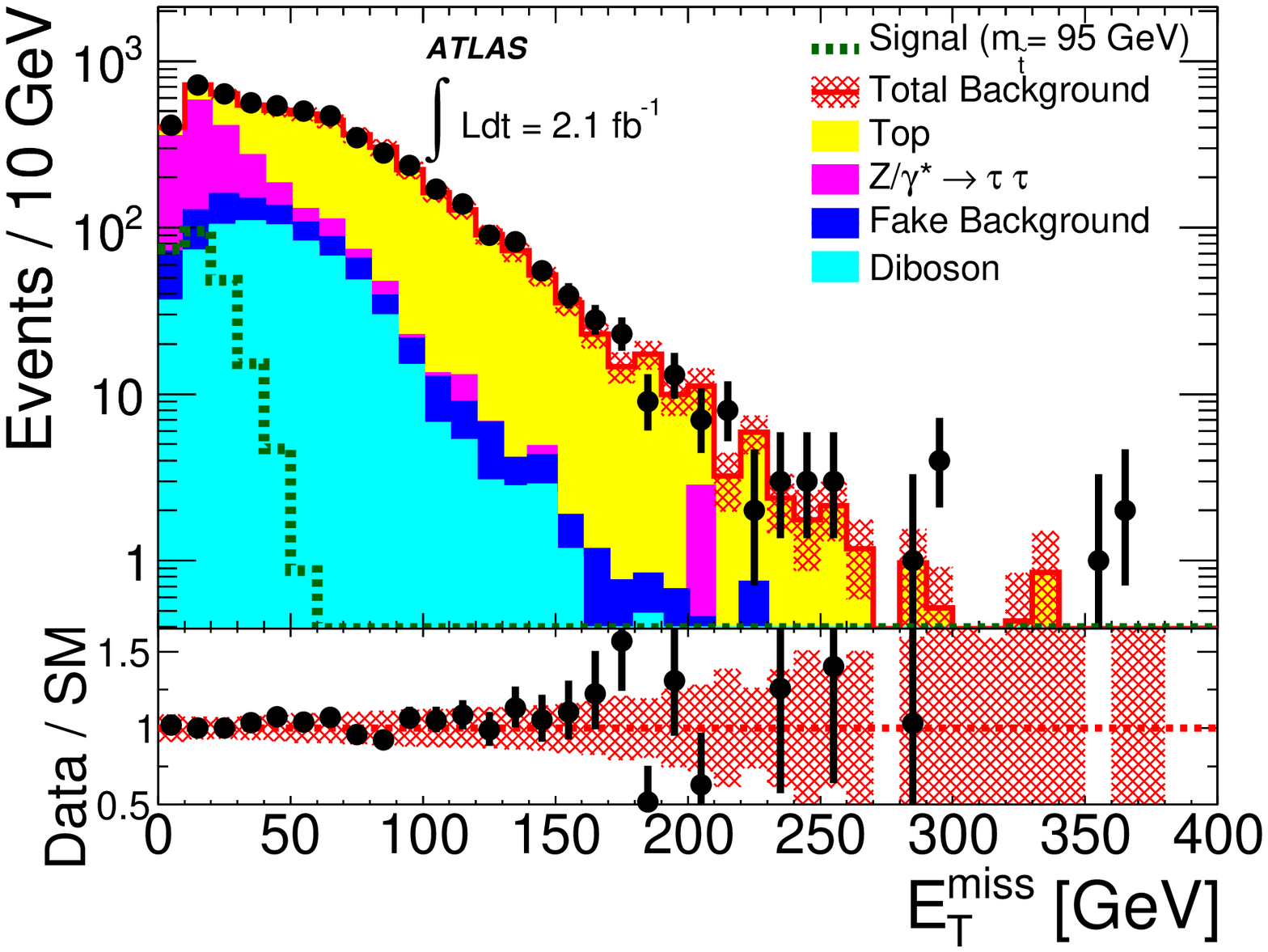}
  \includegraphics[width=0.49\textwidth]{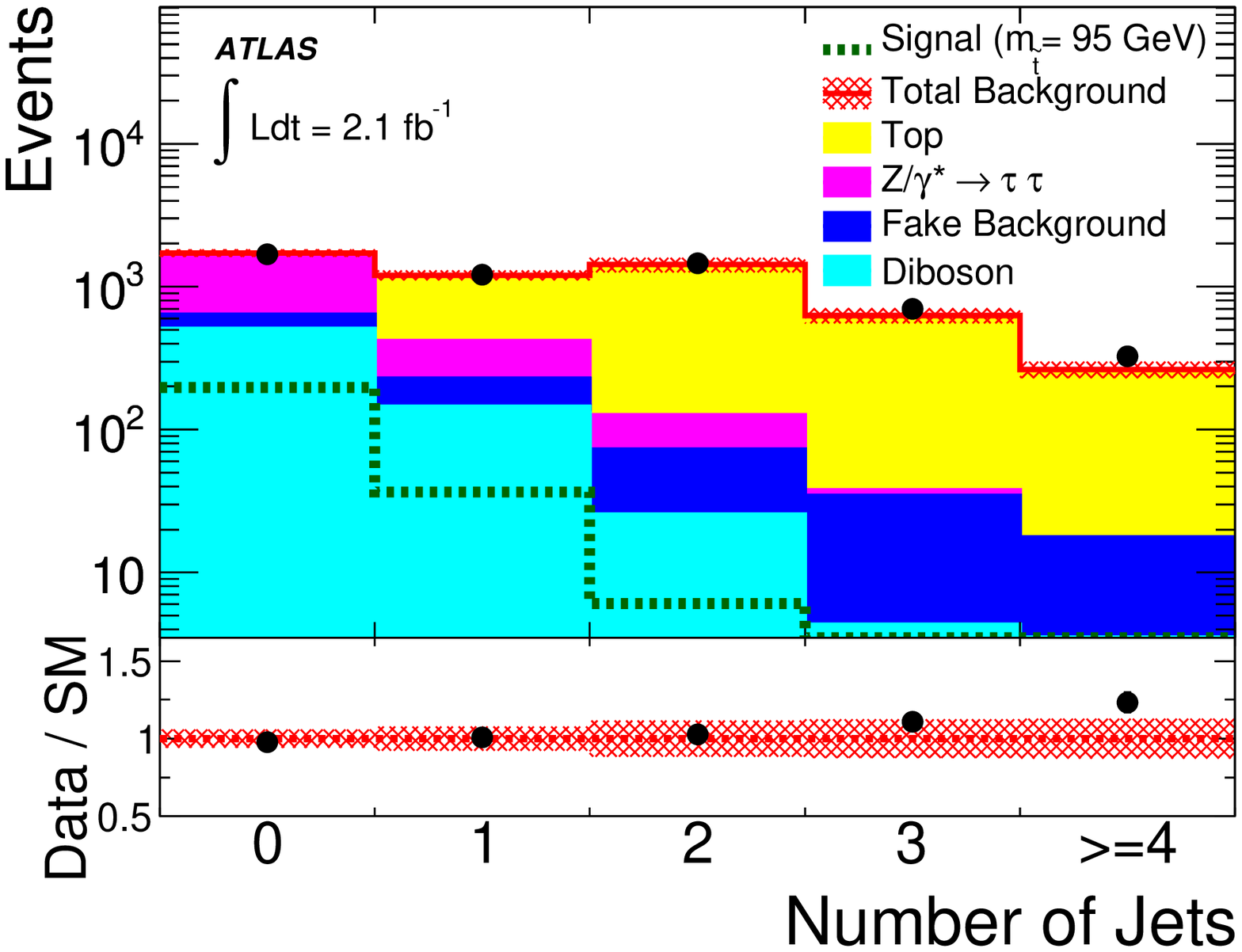}
\end{center}
\caption{Observed distributions of dilepton invariant mass ($m_{e\mu}$), dilepton azimuthal opening angle ($\Delta \phi_{e\mu}$), $\met$ and number of jets after object selection (`preselection').  The expected SM contributions, obtained as described in the text, with combined statistical and systematic uncertainties, are shown.   In addition, the expected signal for  $m_{\tilde{t}}=95\GeV$ is overlaid.  For each case, a plot of the ratio of observed events to the expected background is shown.  The error bars on these points represent the statistical errors on the data points and the hashed boxes represent the total error (statistical and systematic) on the expected background. }
\label{fig:data_mc_preselect}
\end{figure*}

The middle column of Table~\ref{tab:data_mc_cutflow} gives the number of events in the data and the estimated background contributions with their total uncertainties after the event preselection.  A total of 5387 $e\mu$ candidates are observed with 5300 $\pm$ 400 events expected from SM processes.  The number of expected signal events is shown for $m_{\tilde{t}}=$ 95, 250, 500, and 1000~GeV, assuming $|\lambda_{131}^{\prime} \lambda_{231}^{\prime}| = |\lambda_{132}^{\prime}\lambda_{232}^{\prime}| = 0.05$.
Figure~\ref{fig:data_mc_preselect} shows the comparison between data and the expected SM background 
for the dilepton invariant mass ($m_{e\mu}$), their azimuthal opening angle ($\Delta \phi_{e\mu}$), 
$\met$ and the number of jets. 
A good description of the data by the expected SM background is observed.

\begin{table}
\begin{center}
\caption{\label{tab:data_mc_cutflow} Number of events observed in data, the estimated backgrounds, and expected number of signal events, assuming $|\lambda_{131}^{\prime}\lambda_{231}^{\prime}|=|\lambda_{132}^{\prime}\lambda_{232}^{\prime}|=0.05$, with their combined systematic and statistical uncertainties for the preselected sample and the final selected sample.  The number of signal and background events has been rounded.}
\begin{tabular}{ccc}
\hline
\hline
  Process & Preselection &Final selection \\ \hline
$WW$                                                   & 640  $\pm$ 50      & 23.4 $\pm$ 3.3  \\
$Z/\gamma^* \rightarrow \tau \tau$  & 1210 $\pm$ 110   & \ 10  $\pm$ 4  \\
Fake Background                                 & 290  $\pm$ 40      & \ 9.6  $\pm$ 1.9  \\
$WZ$                                                    &36   $\pm$ 4  & \ 0.76 $\pm$ 0.31 \\
$t\bar{t}$                                             & 2800 $\pm$ 400    & \ 0.25 $\pm$ 0.17 \\
Single top                                            & 270  $\pm$ 40       & \ 0.22  $\pm$ 0.20  \\
$W/Z+\gamma$                                 &20   $\pm$ 7   & \ 0.04 $\pm$ 0.04 \\
$ZZ$                                                   &\ 4.0  $\pm$ 0.4    & \ 0.042 $\pm$ 0.028 \\ 
Total background                               &5300 $\pm$ 400     & 44 $\pm$ 6  \\ \hline
Data                                                   & 5387                        & 39              \\ \hline
Signal ($m_{\tilde{t}} = 95$ GeV)       & 240  $\pm$ 15         & 67  $\pm$ 5  \\ 
Signal ($m_{\tilde{t}} = 250$ GeV)     & 23.7  $\pm$ 1.4 & 9.3  $\pm$ 0.6 \\ 
Signal ($m_{\tilde{t}} = 500$ GeV)     &\ 3.05 $\pm$ 0.18   & 1.28 $\pm$ 0.08  \\
Signal ($m_{\tilde{t}} = 1000$ GeV)   &\ 0.305  $\pm$ 0.018  & 0.124  $\pm$ 0.008  \\
\hline
 \hline
\end{tabular}
\end{center}
\end{table}

To increase the signal purity, the preselected events are required to have zero jets,  $\memu >100$~GeV, 
$\Delta \phi_{e\mu}>3.0$ rad and \met~$<25$~GeV.  This selection was optimized using the signal sample with $m_{\tilde{t}} = 95$~GeV which is the most demanding in terms of signal-to-background ratio when setting limits.
After applying the full selection, 39 events are observed with 44 $\pm$ 6 SM events expected.
A breakdown of the SM background composition is given in the last column of Table~\ref{tab:data_mc_cutflow}.  
In order of importance, the dominant contributions stem from $WW$, $\tau$-pair and fake background.  
The $\memu$ distribution of the selected events is shown in Fig.~\ref{fig:emu_mass}.

Systematic uncertainties on the SM background estimation arise from uncertainties in the estimation of the fake background (15\%), the integrated luminosity (3.7\%), and lepton trigger, reconstruction and identification efficiencies (1--2\%).  Uncertainties from lepton energy/momentum scale and resolution (0.5--1\%), \met~modelling (12\%), and jet energy scale and resolution~\cite{jettools} (3.6\%) are also included.  The SM background uncertainty in the shape of the $\memu$ distribution used to extract the signal is estimated by comparing the default $WW$ distribution generated with {\sc herwig}~\cite{herwig} to those obtained with {\sc alpgen}~\cite{alpgen}  (interfaced with {\sc jimmy}~\cite{jimmy}) and {\sc sherpa}~\cite{sherpa}. A 13\% uncertainty is assigned.  The  uncertainties on the $t\bar{t}$ and single-top cross sections are 10\%~\cite{topcs} and 9\%~\cite{singletop}, respectively.  The theoretical uncertainties assigned to the $W/Z+\gamma$, $Z/\gamma^* \rightarrow \tau\tau$, $WW$, $WZ$, and $ZZ$ cross sections are 10\%, 5\%, 7\%, 7\%,  and 5\% respectively; these arise from the choice of PDFs, the factorization and renormalization scale dependence, and $\alpha_s$ variations.

\section{Limit Setting}
\begin{figure}[ht]
\begin{center}
  \includegraphics[width=0.5\textwidth]{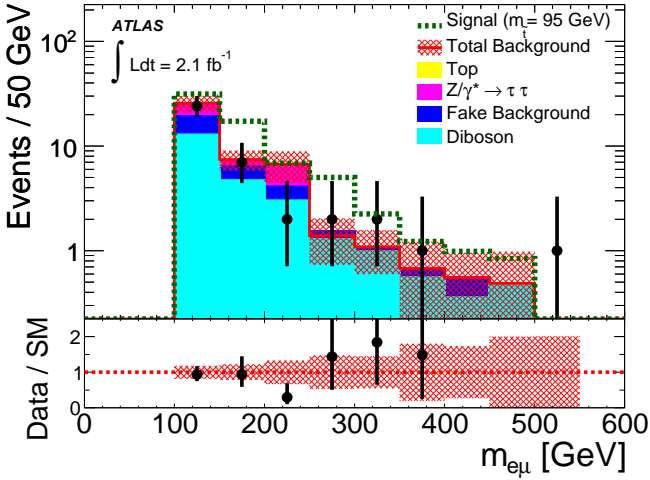}
\end{center}
\caption{The observed $m_{e\mu}$ distribution after applying all selection criteria. The expected SM contributions, obtained as described in the text, with combined statistical and systematic uncertainties, are shown.   In addition, the expected signal for  $m_{\tilde{t}}=95\GeV$ is overlaid.  Finally, a plot of the ratio of observed events to the expected background is shown.  The error bars on these points represent the statistical errors on the data points and the hashed boxes represent the total error (statistical and systematic) on the expected background.}
\label{fig:emu_mass}
\end{figure}

Since no excess is observed in data, the $\memu$ distribution in Figure~\ref{fig:emu_mass}, with a single bin for   $\memu > 400$ GeV to reduce sensitivity to statistical fluctuations, is used to set limits on the production cross section of $e\mu$ pairs through $t$-channel 
exchange of $\tilde{t}$ in RPV SUSY models. A modified frequentist approach, using a binned log-likelihood ratio (LLR) of the signal-plus-background 
hypothesis to the background only hypothesis~\cite{clcalc1}, is used to set the 95\% confidence level (CL) upper limits. 
Confidence levels, $\rm{CL}_{s+b}$ and $\rm{CL}_b$,  are defined by integrating the normalized probability distribution of LLR values 
from the observed LLR value to infinity for the two hypotheses. 
Since no data excess is observed, the production cross section is
excluded at $95\%$ CL when $1-\rm{CL}_{s+b}/\rm{CL}_b=0.95$.  The limits take into account systematic uncertainties by convolving the Poisson probability distributions for signal and background with the probability distributions for the corresponding uncertainty, which are assumed to be Gaussian.

%\FloatBarrier

\begin{figure*}[!hb]
\begin{center}
\subfigure[]{
\label{fig:limitsa}
\includegraphics[width=0.48\textwidth]{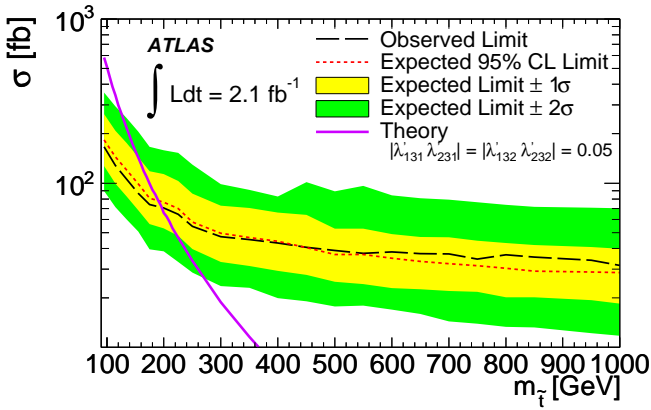}
}
\subfigure[]{
\label{fig:limitsb}
\includegraphics[width=0.48\textwidth]{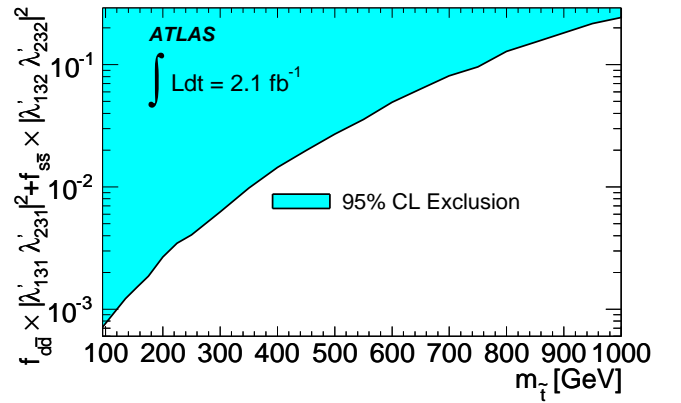}
}
\caption{(a) The observed 95\% CL upper limits on $\sigma(pp \rightarrow e\mu)$ through the $t$-channel exchange of a scalar top quark as a function
of $m_{\tilde{t}}$. The expected limits are also shown together with the $\pm 1$ and $\pm 2$ standard deviation
uncertainty bands. The theoretical cross section for 
$|\lambda_{131}^{\prime}\lambda_{231}^{\prime}|=|\lambda_{132}^{\prime}\lambda_{232}^{\prime}|=0.05$ is also shown. (b) Excluded region for the PDF weighted sum of couplings ($f_{d\bar{d}} \times |\lambda_{131}^{\prime}\lambda_{231}^{\prime}|^2 + f_{s\bar{s}} \times |
\lambda_{132}^{\prime}\lambda_{232}^{\prime}|^2$) as a function of $m_{\tilde{t}}$.
\label{fig:limits}}
\end{center}
\end{figure*}

The upper limit on the production cross section for $pp \rightarrow e\mu X$ through the $t$-channel exchange 
of a $\tilde{t}$ at 95\% CL is shown in Fig.~\ref{fig:limitsa}. 
For a $\tilde{t}$ with mass of 95~GeV (1000~GeV), the limit on the production cross section is 170 (30)~fb which is in agreement with the expected limit of $180_{-60}^{+80}$ ($30_{-10}^{+11}$)~fb. 
The theoretical cross section for $|\lambda_{131}^{\prime}\lambda_{231}^{\prime}|=|\lambda_{132}^{\prime}\lambda_{232}^{\prime}|=0.05$ is also shown to illustrate the sensitivity.

The fraction of events produced by the $d\bar{d}~\rightarrow~e\mu$ ($s\bar{s}~\rightarrow~e\mu$) process is predicted to be $f_{d\bar{d}}=0.72$ ($f_{s\bar{s}}=0.28$) 
using the {\sc pythia} generator with the central CTEQ6L1 PDF set and with $\mstop = 95$~GeV.
The cross section for the signal process is hence proportional to the PDF-weighted sum of the RPV couplings, which is 
$f_{d\bar{d}} \times |\lambda_{131}^{\prime}\lambda_{231}^{\prime}|^2 + f_{s\bar{s}} \times|\lambda_{132}^{\prime}\lambda_{232}^{\prime}|^2$.
The cross section limits set above can be interpreted as a limit on the plane spanned by the sum of couplings
and $m_{\tilde{t}}$. 
The resulting two-dimensional 95\% confidence limit is shown in Fig.~\ref{fig:limitsb}.  

Assuming the equality of all couplings considered in this analysis ($\lambda_{i3j}^{\prime} = \lambda_{131}^{\prime} = \lambda_{231}^{\prime} = \lambda_{132}^{\prime} = \lambda_{232}^{\prime}$), it is possible to compare this result with the one obtained by H1 for masses higher than the centre-of-mass collision energy of 319 GeV available at HERA.  For example, at $m_{\tilde{t}}=400$ (1000) GeV this analysis sets limits on a single coupling, $ \lambda_{i3j}^{\prime}$, of 0.35 (0.70), compared to the limits set by H1 experiment, which are 0.38 (0.95)~\cite{H1}.

\section{Conclusion}
This paper presents a search for LFV interactions in the $e \mu$ continuum, 
as modelled by the $t$-channel exchange of a scalar top quark, using 2.1~fb$^{-1}$ of data collected 
by the ATLAS detector in $\sqrt{s}=7$ TeV $pp$ collisions at the LHC. 
The data are found to be consistent with the SM predictions.
Upper limits are set on the production cross section for $pp \rightarrow e\mu X$ through the $t$-channel exchange 
of a $\tilde{t}$. 
A two dimensional limit in the plane of the weighted sum of couplings vs $m_{\tilde{t}}$ is also obtained.

\section*{Acknowledgements}

We thank CERN for the very successful operation of the LHC, as well as the
support staff from our institutions without whom ATLAS could not be
operated efficiently.

We acknowledge the support of ANPCyT, Argentina; YerPhI, Armenia; ARC,
Australia; BMWF, Austria; ANAS, Azerbaijan; SSTC, Belarus; CNPq and FAPESP,
Brazil; NSERC, NRC and CFI, Canada; CERN; CONICYT, Chile; CAS, MOST and NSFC,
China; COLCIENCIAS, Colombia; MSMT CR, MPO CR and VSC CR, Czech Republic;
DNRF, DNSRC and Lundbeck Foundation, Denmark; EPLANET and ERC, European Union;
IN2P3-CNRS, CEA-DSM/IRFU, France; GNAS, Georgia; BMBF, DFG, HGF, MPG and AvH
Foundation, Germany; GSRT, Greece; ISF, MINERVA, GIF, DIP and Benoziyo Center,
Israel; INFN, Italy; MEXT and JSPS, Japan; CNRST, Morocco; FOM and NWO,
Netherlands; RCN, Norway; MNiSW, Poland; GRICES and FCT, Portugal; MERYS
(MECTS), Romania; MES of Russia and ROSATOM, Russian Federation; JINR; MSTD,
Serbia; MSSR, Slovakia; ARRS and MVZT, Slovenia; DST/NRF, South Africa;
MICINN, Spain; SRC and Wallenberg Foundation, Sweden; SER, SNSF and Cantons of
Bern and Geneva, Switzerland; NSC, Taiwan; TAEK, Turkey; STFC, the Royal
Society and Leverhulme Trust, United Kingdom; DOE and NSF, United States of
America.

The crucial computing support from all WLCG partners is acknowledged
gratefully, in particular from CERN and the ATLAS Tier-1 facilities at
TRIUMF (Canada), NDGF (Denmark, Norway, Sweden), CC-IN2P3 (France),
KIT/GridKA (Germany), INFN-CNAF (Italy), NL-T1 (Netherlands), PIC (Spain),
ASGC (Taiwan), RAL (UK) and BNL (USA) and in the Tier-2 facilities
worldwide.

\clearpage
\onecolumn
% ATLAS Collaboration author list for 13-FEB-2011
% Data extracted on 20-Apr-2012 for paperid 198
%\documentclass[11pt]{article}
%\usepackage{a4wide}\begin{document}
\begin{flushleft}
{\Large The ATLAS Collaboration}

\bigskip

G.~Aad$^{\rm 48}$,
B.~Abbott$^{\rm 111}$,
J.~Abdallah$^{\rm 11}$,
A.A.~Abdelalim$^{\rm 49}$,
A.~Abdesselam$^{\rm 118}$,
O.~Abdinov$^{\rm 10}$,
B.~Abi$^{\rm 112}$,
M.~Abolins$^{\rm 88}$,
H.~Abramowicz$^{\rm 153}$,
H.~Abreu$^{\rm 115}$
% ...
\bigskip

% ...
$^{10}$ Institute of Physics, Azerbaijan Academy of Sciences, Baku, Azerbaijan\\
$^{11}$ Institut de F\'isica d'Altes Energies and Departament de F\'isica de la Universitat Aut\`onoma  de Barcelona and ICREA, Barcelona, Spain\\
% ...
$^{48}$ Fakult\"{a}t f\"{u}r Mathematik und Physik, Albert-Ludwigs-Universit\"{a}t, Freiburg i.Br., Germany\\
$^{49}$ Section de Physique, Universit\'e de Gen\`eve, Geneva, Switzerland\\
% ...
$^{88}$ Department of Physics and Astronomy, Michigan State University, East Lansing MI, United States of America\\
% ...
$^{111}$ Homer L. Dodge Department of Physics and Astronomy, University of Oklahoma, Norman OK, United States of America\\
$^{112}$ Department of Physics, Oklahoma State University, Stillwater OK, United States of America\\
% ...
$^{115}$ LAL, Universit\'e Paris-Sud and CNRS/IN2P3, Orsay, France\\
% ...
$^{118}$ Department of Physics, Oxford University, Oxford, United Kingdom\\
% ...
$^{153}$ Raymond and Beverly Sackler School of Physics and Astronomy, Tel Aviv University, Tel Aviv, Israel\\
% ...
\end{flushleft}

%\end{document}


\begin{thebibliography}{99}
\bibitem{coordinate} ATLAS uses a right-handed coordinate system with its origin at the nominal interaction point (IP) in the centre 
of the detector and the $z$-axis along the beam pipe. The $x$-axis points from the IP to the centre of the LHC ring, and the $y$-axis points upward. Cylindrical coordinates $(r,\phi)$ are used in the transverse plane, $\phi$ being the azimuthal angle around the 
beam pipe. The pseudorapidity is defined in terms of the polar angle $\theta$ as $\eta=-\ln\tan(\theta/2)$.
\end{thebibliography}
\end{document}